# Changing users' security behaviour towards security questions: A game based learning approach

Nicholas Micallef
Australian Centre for Cyber Security
School of Engineering and Information Technology
University of New South Wales
Canberra, Australia
n.micallef@adfa.edu.au

Nalin Asanka Gamagedara Arachchilage
Australian Centre for Cyber Security
School of Engineering and Information Technology
University of New South Wales
Canberra, Australia
nalin.asanka@adfa.edu.au

*Abstract*— Fallback authentication is used to retrieve forgotten passwords. Security questions are one of the main techniques used to conduct fallback authentication. In this paper, we propose a serious game design that uses system-generated security questions with the aim of improving the usability of fallback authentication. For this purpose, we adopted the popular picture-based *"4 Pics 1 word"* mobile game. This game was selected because of its use of pictures and cues, which previous psychology research found to be crucial to aid memorability. This game asks users to pick the word that relates to the given pictures. We then customized this game by adding features which help maximize the following memory retrieval skills: (a) verbal cues - by providing hints with verbal descriptions; (b) spatial cues - by maintaining the same order of pictures; (c) graphical cues - by showing 4 images for each challenge; (d) interactivity/engaging nature of the game.

## I. Introduction

Republican vice presidential candidate Sarah Palin's Yahoo! email account was "hijacked" in the run-up to the 2008 US election. The "hacker" simply used the password reset prompt and answered her security questions [1]. As reported [1], the Palin hack didn't require much technical skills. Instead, the hacker merely used social engineering techniques to reset Palin's password using her birthdate, ZIP code and information about where she met her spouse. The answers to these questions were easily accessible with a quick Google search. Also, as more of our personal information is available online, it is becoming easier for attackers to retrieve this information, through observational attacks, from social networking websites, such as Facebook [2], Twitter or even more professional websites like LinkedIn [3]. Besides observational attacks, security questions are also vulnerable to guessing attacks, in which, attackers try to access accounts by providing low entropy (i.e., level of complexity) answers (e.g., favorite color: blue). These attacks are part of a series of Cyber-threats which usually include computer viruses and other types of malicious software (malware), unsolicited e-mail (spam), eavesdropping software (spyware), orchestrated campaigns aiming to make computer resources unavailable to the intended users (distributed denial-of-service (DDoS) attacks), social engineering, and online identity theft (phishing). Hence, the ease of conducting observational and guessing attacks has increased the vulnerabilities of fallback authentication mechanisms [4] towards all these cyber-threats, which are leading to severe consequences, such as monetary loss, embarrassment and inconvenience [5].

A possible way to reduce the vulnerability of security questions towards these kind of attacks is by encouraging users to use system-generated answers [5]. One particular technique uses an *Avatar* to represent system-generated data of a fictitious person (see Figure 1), and then the Avatar's system-generated data is used to answer security questions [5]. However, the main barrier towards widespread adoption of these techniques is memorability [6], since users struggle to remember the details of system-generated information to answer their security questions.

Thus, to address this problem with memorability of system-generated data, in this paper we present a game design that focuses on enhancing users' memorability of answers to security questions. This paper investigates the elements (obtained from the literature [7] [8] [9] [10]) that should be addressed in the game design to create and consequently nurture the bond between users and their avatar profiles (system-generated data). For the purpose of our research, we adopted the popular picture-based *"4 Pics 1 Word"* [1] mobile game. This game asks users to pick the word that relates the given pictures (e.g., for the pictures in Figure 2a the relating word would be "Germany"). This game was selected because of its use of pictures and cues, in which, previous psychology research has found to be important to help with memorability [7] [11].

For the purpose of our research we adopted the game, so that at certain intervals, it asks users to solve avatar-based challenges. Since previous research on memorability found that recognition is a simpler memory task than recall [12], besides recall-based challenges (see Figure 3a), in our game, we also provide recognition-based challenges (see Figure 3b). Hence, the proposed game design focuses on encoding the system-generated data to users' long-term memory [11] and to aide memorability by using the following memory retrieval skills [13]: (a) graphical cues - by using images in each challenge; (b) verbal cues - by using verbal descriptions as hints; (c) spatial cues - by keeping same order of pictures; and (d) interactivity - interactive/engaging nature of the game through the use of persuasive technology principles [9].

---

[1] https://play.google.com/store/apps/details?id=de.lotum.whatsinthefoto.us&hl=en

In the following sections, we describe the fallback authentication mechanisms that are currently being used. We then identify the strengths and weaknesses of research on security questions to show why our research is important and how it is considerably different from previous research that has been conducted in this field. Afterwards, we describe the main contribution of this paper, which is a unique game design that uses gamification and memorability concepts to improve the memorability of fallback authentication. Finally, we conclude this paper by presenting the prototype that we will use to evaluate the proposed game design in a lab study.

## II. BACKGROUND

As computer users have to deal with an increasing number of online accounts [14] [15] they are finding it more difficult to remember all passwords for their different accounts. For example, if we look just at social networking websites, plenty of users have different accounts for Facebook, Twitter, Instagram, SnapChat and LinkedIn. Since password managers have not been widely adopted [16], resetting of passwords is becoming a more frequent task [14] [15]. To address this problem various forms of fallback authentication mechanisms have been evaluated with the most popular being security questions [17] (focus of this research) and email-based password reset. Although email-based (or in some cases even SMS-based) password recovery has been widely adopted by major organizations (e.g., Google) they still have the limitation of being vulnerable to 'man in the middle' attacks, since these emails are not encrypted [18]. Other fallback authentication mechanisms (e.g., social authentication [19]) have also been evaluated though they have not been widely adopted [20], since they are vulnerable to impersonation both by insiders and by face-recognition tools [21].

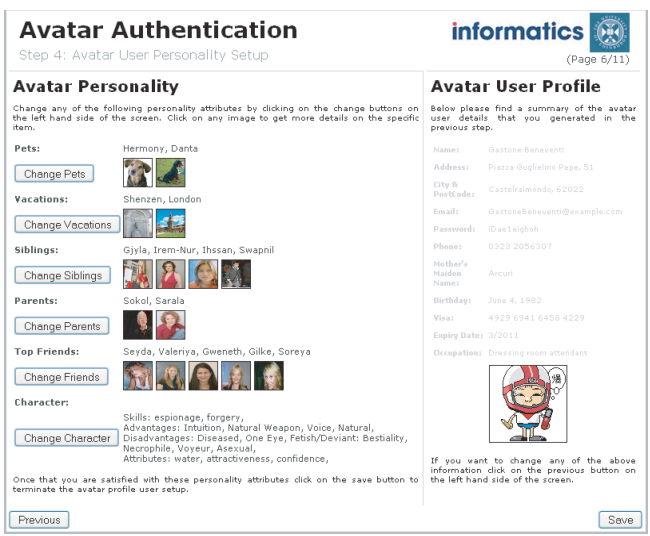

Figure 1. System-generated Avatar profile as defined by Micallef and Just 2011 [5]

Security questions are the most widely adopted form of fallback authentication [20] [15] since they are used by a variety of popular organizations (e.g., Banks, E-commerce websites, Social networks). Security questions are set-up at account creation. Then when they want to reset their password, users will have to recall the answers that they provided when setting up the account. Several studies have found that security questions have the following major limitations: (1) can be guessed by choosing the most popular answers [3]; (2) have memorability problems since they are not frequently used [6], which decreases their level of usability [22]; (3) are easily guessed by friends, family members and acquaintances [23] [24]; (4) can be guessed by observational attacks, with a quick Google search or by searching victims' social networking websites [2]. Recent studies, conducted using security questions data collected by Google [22], found that security questions are neither usable (low memorability) nor secure enough to be used as the main account recovery mechanism. This means that new techniques need to be investigated to provide a more secure and memorable form of fallback authentication.

In the last years, mobile devices became one of the main mediums to access the web and people started storing (and accessing) more sensitive information on these devices [25]. Hence, the focus of authentication research has shifted to primarily investigate new techniques (e.g., data driven authentication using sensors [26]) to conduct authentication on mobile devices [27] [28]. Most of the research in this area tried to leverage the use of the variety of inbuilt sensors (e.g., accelerometer, magnetometer) that are available on today's mobile devices, with the main goal of striking a balance between usability and security when conducting authentication [29] [30]. However, sensors have also been used in fallback authentication mechanisms on smartphones [31] as a technique that extracts autobiographical information [32] about the users' smartphone behavior during the last couple of days. This information is then used to answer security questions about recent smartphone use [33]. Although these innovative security questions techniques have managed to achieve memorability rates of about 95% using a diverse set of questions [34] [35], these techniques have mostly been evaluated with a younger user-base (mean age of 26), those users that use smartphones the most [36]. Hence, we argue that other techniques need to be investigated to cater for those users who do not use smartphones or use them but not very frequently (e.g. age 50+).

Besides the previously described work on autobiographical security questions, recent research has also investigated: (1) life-experiences passwords - which consists of several facts about a user-chosen past experience, such as a trip, graduation, or wedding, etc. [37]; (2) security meters - to encourage users to improve the strength of their security answers [38] and (3) avatar profiles - to represent system-generated data of a fictitious person (see Figure 1), and then the Avatar's information is used to answer security questions [5]. Although life-experience passwords [37] were evaluated to be stronger then passwords and less guessable then security questions. However, the memorability after 6 months was still about 50%. The work on security meters for security questions [38] seems to be quite promising, however it is still at an embryonic stage and it requires further research to evaluate its feasibility.

Using system-generated data (see Figure 1), in the form of an avatar profile, to answer security questions [5] has also not been extensively investigated. However, in our research we attempt to investigate this work further because compared to other research on security questions it seems to be the one that has the potential to achieve the optimal balance in terms of security and memorability due to the following reasons: (1) it could be tailored for everyone (and not only for those users with medium/high smartphone usage); (2) guessing attacks could be minimized because the entropy and variety of the answers could be defined/controlled by the system that generates them; (3) risks of having observational attacks would be minimal since the system-generated avatar information would not be publicly available; and (4) memorability could be achieved by using a gamified approach to create and nurture a bond between users and their avatar profiles (in the form of system-generated data as in Figure 1).

Bonneau and Schechter found that most users can memorize passwords when using tools that support learning over time [39]. However, we know to our cost, no-one has attempted to use serious games to improve the users' memorability of systems-generated answers for security questions. Thus, in our research, we attempt to use a gamified approach to improve users' memorability during fallback authentication because previous work in the security field [40] has successfully used this approach to educate users about the susceptibility to phishing attacks [41] with the aim of teaching users to be less prone to these types of security vulnerabilities [42]. Hence, this paper contributes to the field of fallback authentication by proposing a game design which uses long-term memory and memory retrieval skills [13] to improve the memorability of security answers based on a system-generated avatar profile.

## III. GAME DESIGN

The main challenge in designing usable security questions mechanisms is to create associations with answers that are strong and to maintain them over time. In our research we use previous findings on the understanding of long-term memory to design a game which has the aim of improving the memorability of system-generated answers for security questions. Atkinson and Shiffrin [11] proposed a cognitive memory model, in which, new information is transferred to short-term memory through the sensory organs. The short-term memory holds this new information as mental representations of selected parts of the information. This information is only passed from short-term memory to long-term memory when it can be encoded through cue-association [11] (e.g., when we see a cat it reminds us of our first cat). This encoding through cue-association helps people to remember and retrieve the stored information over an extended period of time. These encodings are strengthened through constant rehearsals. Also, psychology research has found that humans are better at remembering images than textual information (known as the picture superiority effect) [7]. In section IIIA, we describe how we use these psychology concepts to adopt the popular *"4 Pics 1 Word"* mobile game for the purpose of our research. We create encoding associations (bond) with the avatar profile by using the picture-based nature of this game and by adding verbal cues. Then in section IIIB we describe how we strengthen these encodings by having users constantly rehearse associations (nurture the bond) through persuasive technology principles [9].

### A. Game Features

In most instances, the game functions similarly to the *"4 Pics 1 Word"* mobile game, meaning that the game asks players to pick the word that relates the given pictures (e.g., for the pictures in Figure 2a the relating word would be "Germany"). However, at certain intervals, the game asks players to solve avatar-based challenges. The optimal number of times that players will be given avatar-based challenges during a day to learn the system-generated avatar information will be investigated in a field study. The game provides players with a pool of 12 letters to assist them with solving the challenge. For each given answer, players are either rewarded or deducted points based on whether they provided the correct or wrong answer (10 points when answering standard challenges, 15 points when answering avatar-based recognition challenges, 20 points when answering avatar-based recall challenges). Points can be used to obtain hints to help in solving more difficult challenges (deduction of 30/50 points).

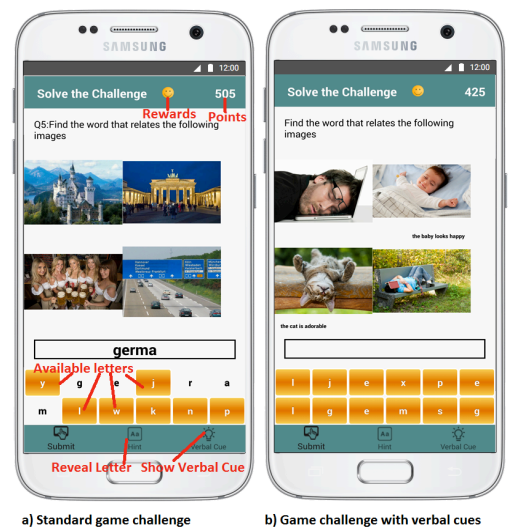

Figure 2. Examples of standard game challenges.

Researchers in psychology have defined two main theories to explain how humans handle recall and recognition: Generate-recognize theory [43] and Strength theory [12]. According to the generate-recognize theory [43] recall is a two phase process: Phase 1 - A list of possible words is formed by looking into long-term memory; Phase 2 - The list of possible words is evaluated to determine if the word that is being looked for is within the list. According to this theory recognition does not use the first phase, hence it's easier and faster to perform. According to strength theory [12] recall and recognition require the same memory tasks, however recognition is easier since it requires a lower level of strength. When it comes to avatar-based challenges, in our game we decided to use both recall and recognition challenges (see Figure 3) because having only recognition challenges would have lowered the security level of the game, since the answer space would have been very small. Hence, to try and strike a balance between security and memorability, we designed the avatar challenges part of the game so that it starts by showing mostly recognition-based challenges (see Figure 3b). Then as players get more accustomed to the avatar profile and they learn the system-generated data (strengthening of the bond) the avatar-based challenges would become mainly recall-based (see Figure 3a).

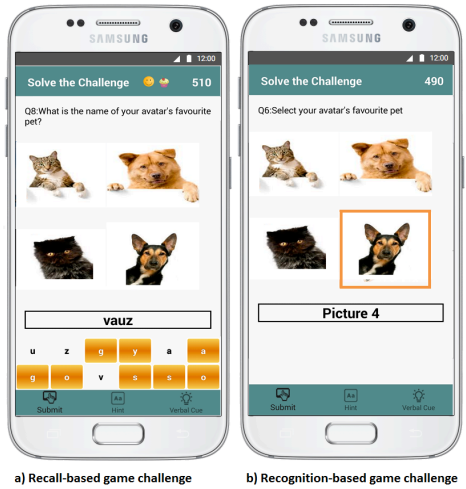

Figure 3. Examples of recall and recognition-based avatar challenges.

Psychology research [43] [44] has shown that it is difficult to remember information spontaneously without having any kind of memory cues. Hence, we added a feature that shows verbal cues about each picture (see Figure 2b). This feature can be enabled by using the points (30/50 points) that are gathered when solving other game challenges as the player goes through the game. We decided to add this feature, especially for the avatar-based challenges, so that players can focus their attention on associating the words with the corresponding cues (pictures). We hypothesize that this should help to process and encode the information in memory and store it in the long-term memory [13].

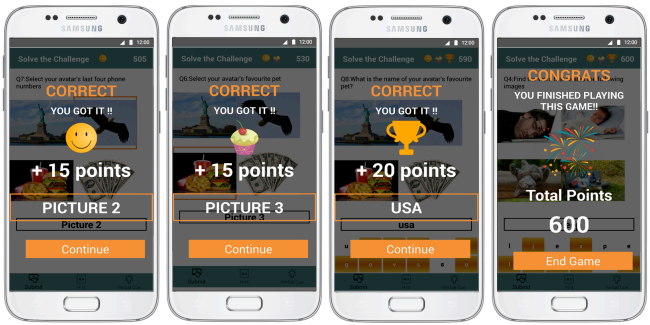

Figure 4. Examples of rewards and game visualizations.

We decided to have a fixed set of images and always show the same images in the same order because this helps enhancing semantic priming [13]. Meaning that it will help players recognize the answer by associating it with the other images that are presented with it. To improve the security element of the game, especially when solving avatar-based challenges, our game does not show the length of the word that needs to be guessed. This feature makes the game more difficult, but we argue that it increases the level of security.

### B. Engagement

To nurture the bond between players and their avatars, we will use constant rehearsals to strengthen the encodings of associations with the system-generated data, in the players' long-term memory. We plan to achieve this by using the following persuasive technology principles proposed by Fogg [9] and also used in [45]:

**Tunnelling:** Tunnelling is the process of providing a game experience which contains opportunity for persuasion [9]. Players are more likely to engage in a tunnelling experience when they can see tangible results [45]. For this reason, at the beginning of the game, the avatar-based challenges are mostly recognition-based rather than recall-based. We hypothesize that in this way it is less likely that players will stop playing the game due to being exposed to difficult challenges at the beginning. Also, at this stage of the game obtaining hints requires a low amount of points (30 points). Additional levels of difficulty (recall-based challenges) become available only as players either demonstrate sufficient skill, or play the game for several days or weeks. As the player goes through the game the cost (in points) of buying hints or obtain verbal cues will increase as well (50 points).

**Conditioning:** According to persuasive technology principles [9] players can be conditioned to play a game if they are offered rewards to compensate their progress. In our game we reward players with points when they solve challenges correctly (more points are given when avatar-based challenges are solved, recall-based challenges provide more points than recognition-based challenges). The more points players collect the more hints they can obtain when they are struggling to solve other game challenges. We also reward players with the following badges (see Figure 4) each time that they solve avatar-based challenges: (1) a "smiley" badge when they solve 1 avatar challenge (see Figure 4a); (2) a "cake" badge when they solve half of the daily avatar challenges (see Figure 4b); (3) a "trophy" badge when they solve all daily avatar challenges (see Figure 4c). Special sounds and visualizations are displayed when these badges or an important milestone is achieved (see Figure 4d).

**Suggestion:** Persuasive technology principles [9] suggest that messages and notifications should be well timed in order to be more effective. For this reason in our game we send notifications to remind players to play the game every 24 hours, if they did not play the game during that time frame. Also, every 24 hours we provide hints when players are stuck with a game challenge.

**Self-monitoring:** Persuasive technology principles [9] state that constantly showing progress can motivate players to improve their performance. For this reason, in our game we show the score and the progress in solving avatar-based challenges each time that players play the game. We also show graphs on how many avatar-based challenges were solved correctly during a day/week/month and how many challenges still need to be solved to progress to the next stage. We hypothesize that these tools will help players identify areas for improvement and provide motivation to continue playing the game with the aim of improving performance.

**Surveillance and Social Cues:** According to persuasive technology [9], players are more encouraged to perform certain actions if others are aware of these actions and by leveraging social cues. In our game, we implement a social element of surveillance by: (1) congratulating players when they return to play the game every day; (2) applaud players when they reach an important game milestone; (3) encourage players even when they get incorrect answers; (4) express disappointment when players don't play the game regularly.

**Humour, Fun and Challenges:** Affect is also an important factor to enhance players' motivation [45]. To make the game more fun we included emoticons when sending reminders or when communicating with players. This is also the reason why we selected humoristic badges (smiley, cake, trophy) to reward players when they reach avatar-related milestones (see Figure 4). Our motivation is to keep players interested and engaged in playing the game.

## IV. PROTOTYPE GAME LOGIC

In our lab study we plan to evaluate a game prototype by using the following logic. As shown in Figure 5, the game starts by picking a random standard challenge from a pool of 7 standard challenges (all players will experience the same standard challenges but in a random order). After completing a standard challenge, the game player is deducted/awarded points. Afterwards, the challenge is removed from the pool of available challenges. At this stage the player is presented with a randomly selected avatar-based recognition challenge (based on the avatar profile that they selected prior to playing the game). If the player picks the correct answer, a badge is rewarded based on how many avatar-based challenges they solved. The player will continue to be presented with alternate standard and avatar-based recognition challenges until they complete the 3 avatar-based recognition challenges. After that, the player is prompted with alternate standard and avatar-based recall challenges until all 3 recall avatar-based challenges are completed. This is where the game ends. In total, each player will complete 7 standard challenges, 3 recognition and 3 recall avatar-based challenges.

## V. CONCLUSIONS AND FUTURE WORK

The proposed game design outlined in this paper teaches and nudges users to provide stronger answers to security questions to protect themselves against observational and guessing attacks. Since this technique uses system-generated data (see Figure 1), it is quite unlikely that attackers would be able to retrieve the avatar-based answers from google searches/social networks or through guessing attacks. We believe that helping users to memorize the avatar's system-generated data through an engaging/interactive gamified approach can help users create and nurture a bond with their avatar. This will be achieved by encoding information in long-term memory through constant rehearsals with the aim of improving memorability of fallback authentication (i.e., security questions). In our future work, we will conduct studies to involve users in this game design (by using the prototype described in section IV and logic shown in Figure 5) to further optimize the functionalities of the game and determine any security vulnerabilities that need to be addressed. Afterwards, we will conduct a field study to evaluate different rehearsal configurations to determine the optimal rehearsal parameters that need to be adopted so that users could use this game to learn and remember the system-generated avatar profile.

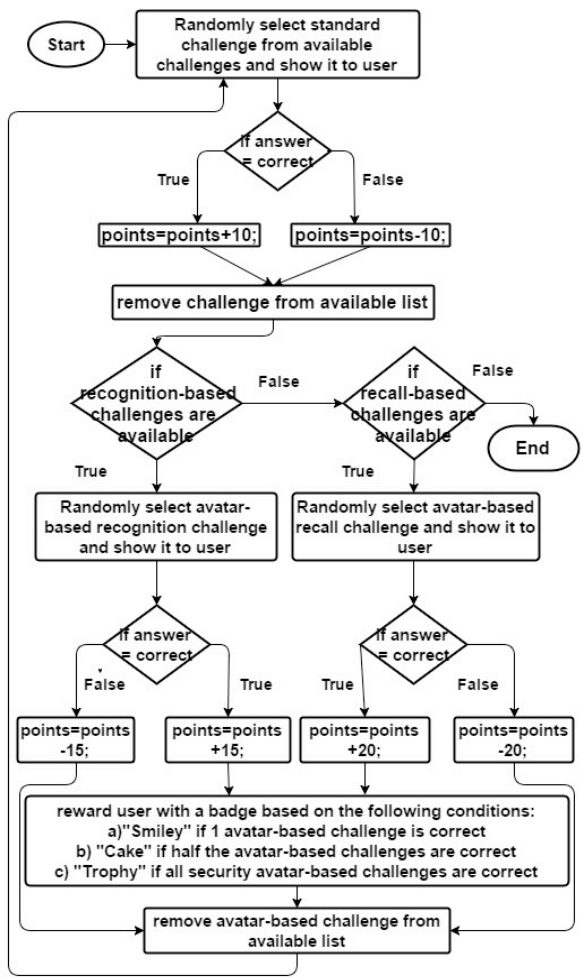

Figure 5. Prototype Game Logic.